\PassOptionsToPackage{usenames,dvipsnames}{color}

\documentclass[10pt,letterpaper,year=24,startpage=1]{fmcad}

\usepackage[T1]{fontenc}
\usepackage[utf8]{inputenc}
\usepackage[english]{babel}
\usepackage{amsmath,amssymb}

\usepackage{todonotes}
\usepackage{microtype}
\usepackage{bussproofs}
\usepackage{dsfont}
\usepackage{listings}
\usepackage{hhline}
\usepackage{subcaption}
\usepackage{mathtools}
\usepackage{booktabs}
\usepackage{tabularx}
\usepackage{multirow}
\usepackage{rotating}
\usepackage{makecell}
\usepackage{cite}

\usepackage[linesnumbered,ruled,vlined]{algorithm2e}
\SetKwBlock{Loop}{loop}{end loop}
\SetKwComment{Comment}{\hfill$\triangleright$~}{}

\SetCommentSty{mycommfont}
\SetKwInOut{Input}{Input}
\SetKwInOut{Output}{Output}
\SetKw{Continue}{continue}

\usepackage{color}
\usepackage{hyperref}
\usepackage[capitalise]{cleveref}

\usepackage[inline]{enumitem}
\setlist{nosep}

\usepackage{prftree}
\usepackage{tikz}
\usetikzlibrary{arrows,calc,math,shapes,decorations,positioning,svg.path}
\tikzset{%
    fontscale/.style = {font=\relsize{#1}},%
}

\definecolor{darkblue}{HTML}{2D2F92}
\definecolor{darkgreen}{rgb}{0,0.5,0}
\definecolor{darkred}{rgb}{0.5,0,0}

\makeatletter
\DeclareRobustCommand{\perc}{%
  \mathbin{\mathpalette\scaletoplus@{\mkern1mu\%\mkern1mu}}%
}
\newcommand{\scaletoplus@}[2]{%
  \vcenter{%
    \hbox{\sbox\z@{$\m@th#1+$}\resizebox{\wd\z@}{!}{$\m@th#1#2$}}%
  }%
}
\makeatother

\newtheorem{example}{Example}
\newtheorem{lemma}{Lemma}

\makeatletter
\def\paragraph{%
    \@startsection{paragraph}{4}%
    {\z@}%
    {0ex plus 0.1ex minus 0.1ex}%
    {0ex}%
    {\normalfont\normalsize\bfseries}%
}%
\def\@IEEEsectpunct{.\ }%
\makeatother

\newcommand\Tstrut{\rule{0pt}{2.2ex}}         

\newcommand{\polysat}{\textsc{PolySAT}}
\newcommand{\stp}{\textsc{STP}}
\newcommand{\cvcfive}{\textsc{cvc5}}
\newcommand{\boolector}{\textsc{Boolector}}
\newcommand{\bitwuzla}{\textsc{Bitwuzla}}
\newcommand{\uclid}{\textsc{uclid}}
\newcommand{\mathsat}{\textsc{MathSAT}}

\newcommand{\yicestwo}{\textsc{Yices2}}
\newcommand{\zthree}{\textsc{Z3}}

\newcommand{\sat}{\texttt{sat}\xspace}
\newcommand{\unsat}{\texttt{unsat}\xspace}

\newcommand{\concat}{\mathbin{+\!\!\!+}}
\newcommand{\bvshl}{\mathbin{<\!\!\!<}}
\newcommand{\bvlshr}{\mathbin{>\!\!\!>}}
\newcommand{\bvashr}{\mathbin{>\!\!\!>}_\mathsf{a}}
\newcommand{\bvand}{\mathbin{\&}}
\newcommand{\bvor}{\mathbin{|}}
\newcommand{\bvnot}{{\sim}}
\newcommand{\bvudiv}{\mathbin{/}}
\newcommand{\bvurem}{\mathbin{\perc}}
\newcommand{\bvule}{\leq_\mathsf{u}}
\newcommand{\bvuge}{\geq_\mathsf{u}}
\newcommand{\bvult}{<_\mathsf{u}}
\newcommand{\bvugt}{>_\mathsf{u}}
\newcommand{\bvsle}{\leq_\mathsf{s}}

\newcommand{\band}[2]{#1 \bvand #2}
\newcommand{\bor}[2]{#1 \bvor #2}
\newcommand{\bnot}[1]{{\bvnot}#1}
\newcommand{\bshl}[2]{#1 \bvshl #2}
\newcommand{\blshr}[2]{#1 \bvlshr #2}
\newcommand{\bashr}[2]{#1 \bvashr #2}
\newcommand{\budiv}[2]{#1 \bvudiv #2}
\newcommand{\burem}[2]{#1 \bvurem #2}

\newcommand{\overmul}[2]{\Omega^*(#1,#2)}

\newcommand{\overadd}[2]{\Omega^+(#1,#2)}

\newcommand{\bvsize}[1]{\lvert #1 \rvert}

\newcommand{\bit}[2]{#1[#2]}
\newcommand{\parity}{\operatorname{parity}}
\newcommand{\parityOf}[1]{\parity(#1)}

\newcommand{\modop}{\operatorname{mod}}
\newcommand{\extract}[3]{#1[#2{:}#3]}   
\newcommand{\Msb}[1]{\operatorname{msb}(#1)}

\newcommand{\roundup}[1]{\lceil #1 \rceil}
\newcommand{\rounddown}[1]{\lfloor #1 \rfloor}
\newcommand{\bigroundup}[1]{\Bigl\lceil #1 \Bigr\rceil}

\newcommand{\eval}[1]{\widehat{#1}}

\newcommand{\interval}[2]{\ensuremath{\mathopen[{#1};{#2}\mathclose[}}
\newcommand{\intervalcc}[2]{\ensuremath{\mathopen[{#1};{#2}\mathclose]}}
\newcommand{\ilen}[1]{\mathit{len}({#1})}

\newcommand{\Zmod}[1]{\mathbb{Z}/{#1}\mathbb{Z}}

\newcommand{\limpl}{\,\rightarrow\,}

\newcommand{\union}{\cup}

\newcommand{\Land}{\bigwedge}

\newcolumntype{A}{ >{$} r <{$} @{} >{${}} l <{$} } 

\newcommand{\fiset}{\mathcal{I}}   
\newcommand{\fijust}{\mathcal{J}}  

\newcommand{\emptyseq}{\langle\rangle}
\newcommand{\pushseq}[2]{\langle{#1};{#2}\rangle}
\newcommand{\fnForward}{\mathit{forward}}
\newcommand{\fnIsConflict}{\mathit{isConflict}}

\newcommand{\fnComputeInterval}{\mathit{computeInterval}}

\title{%
    \polysat{}:
    Word-level Bit-vector Reasoning in Z3
}

\author{%
    \IEEEauthorblockN{%
        Jakob Rath\IEEEauthorrefmark{1} \orcid{0000-0003-0346-6749},
        Clemens Eisenhofer\IEEEauthorrefmark{1} \orcid{0000-0003-0339-1580},
        Daniela Kaufmann\IEEEauthorrefmark{1} \orcid{0000-0002-5645-0292},
        Nikolaj Bj{\o}rner\IEEEauthorrefmark{2} \orcid{0000-0002-1695-2810},
        Laura Kovács\IEEEauthorrefmark{1} \orcid{0000-0002-8299-2714}%
    }%
    \IEEEauthorblockA{%
        \IEEEauthorrefmark{1}%
        TU Wien, Vienna, Austria%
        \quad%
        \IEEEauthorrefmark{2}%
        Microsoft Research, Redmond, USA%
    }%
}

\begin{document}
\maketitle
\begin{abstract}
\polysat{} is a word-level decision procedure supporting bit-precise
SMT reasoning over polynomial arithmetic with large bit-vector operations.
The \polysat{} calculus
extends conflict-driven clause learning modulo theories
with two key components: (i) a bit-vector plugin to the equality graph, 
and (ii) a theory solver for bit-vector arithmetic with non-linear polynomials.
\polysat{} implements dedicated procedures to extract bit-vector intervals from polynomial
inequalities. 
For the purpose of conflict analysis and resolution, \polysat{} comes with on-demand lemma generation over non-linear bit-vector arithmetic. 
\polysat{} is integrated into the SMT solver Z3 and has potential applications
in model checking and smart contract verification where bit-blasting techniques
on multipliers/divisions
do not scale.

\end{abstract}
\section{Introduction\label{sec:introduction}}

Bit-vector reasoning plays a central role in applications of system verification,
enabling for example efficient bounded model checking~\cite{CBMC-ckl2004},
bit-precise memory handling~\cite{Alive2}, or proving safety of decentralized financial transactions~\cite{DBLP:journals/pacmpl/AlbertGRRRS20}. 
Although one may argue that,  because bit-vectors are bounded,
bit-vector reasoning is simpler than proving
arithmetic properties over the integers or reals, showing
(un)satisfiability of bit-vector problems is inherently expensive due to
complex arithmetic operations over large
bit-widths~\cite{DBLP:journals/mst/KovasznaiFB16}.


\paragraph*{Related works} State-of-the-art satisfiability modulo theories (SMT) solvers
handle bit-vector operations by \emph{bit-blasting}~\cite{DBLP:series/txtcs/KroeningS08},
i.e., translating bit-vector formulas into propositional ones  that can be solved
by ordinary propositional satisfiability (SAT) solvers. 
While the core idea of translating bit-vector operations to  SAT formulas
is quite natural,
several variants of such translations arose. 
Some methods apply heavy preprocessing before bit-blasting, see \stp{}~\cite{STP}, whereas others use over- and under-approximations to simplify solving, such as \boolector{}~\cite{Boolector3} and \uclid{}~\cite{UCLID}.
Alternatively, other  approaches bit-blast only relevant parts of the input, as developed in \mathsat{}~\cite{MathSAT} and \cvcfive{}~\cite{cvc5,cvc5LazyBB}.

Yet, the bit-blasting
strategy performs poorly when multiplications are involved.
As a result, stochastic local search, as in \bitwuzla{}~\cite{Bitwuzla} or \zthree{}~\cite{Z3SLS}, 
and int-blasting, as in  \cvcfive{}~\cite{IntBlasting}, have also been developed.
Local search works very well for satisfiable instances, but in general does not terminate for unsatisfiable (unsat) problems.
On the contrary, int-blasting tends to work better  for unsat formulas.


\paragraph*{Our contribution -- \polysat{}} In this paper, we propose \polysat{}, a \emph{word-level reasoning procedure}
as a theory solver integrated into SMT solving. \polysat{} is based on conflict-driven clause learning modulo theories (CDCL(T)), providing thus  \emph{an alternative to bit-blasting}.
Our work builds on and extends previous research on
bit-vector slicing~\cite{DBLP:conf/iccad/BruttomessoS09},
forbidden intervals~\cite{BitvectorsMCSAT},
and fixing bits~\cite{Zeljic-SAT16}.

In our setting, we consider bit-vectors as elements of the ring~$\Zmod{2^w}$. Informally, arithmetical operations on bit-vectors can be seen as the respective integer operations,
where the result is evaluated ``$\modop 2^w$''.
Yet, due to modulo/bounded arithmetic, many properties of the integers
(such as, there is no maximal element and no zero-divisors)
do not hold over bit-vectors.
Nevertheless, with \polysat{} we support bit-vector arithmetic without bit-blasting.

\begin{example} Let us illustrate the benefits of \polysat{} using the following bit-vector constraints with large bit-width~$w$:
\begin{equation*}
  \begin{split}
     xy + y &\bvugt y + 3\\
    6 &= 2y + z
  \end{split}\qquad
  \begin{split}
     1 &= 3x + 6yz + 3z^2\\
    0 &= \band{(2y + 1)}{x}
  \end{split}
\end{equation*}
    where ``$\bvand$'' denotes the bit-wise \emph{and} operation  and $\bvugt$ refers to unsigned comparison.
    \polysat{} proves this set of bit-vector constraints to be  \unsat, without using bit-blasting  as follows. 

    We guess the assignment $x = 0$, simplifying the first constraint to $y \bvugt y + 3$.
    We pick the assignment $y = 2^w-2$ which is feasible w.r.t. the inequality.
    Hence, the constraint $6 = 2y + z$ simplifies to $z = 10$, which conflicts with the constraint $1 = 3x + 6yz + 3z^2$.
    We backtrack, apply variable elimination upon~$y$ on the two equality constraints, and learn the equation $3x + 18z = 1$.
    From the bit-wise $\bvand$-constraint, we derive that~$x$ is even, as $2y + 1$ is odd.
    This, however, conflicts with the learned clause, as it implies that $x$ is odd.
    Hence, \polysat{} concludes that the given constraints are \unsat.
\end{example}

\paragraph*{\polysat{} -- Main improvements}
With \polysat{}, we bring the following main improvements to word-level reasoning over bit-vectors. 
\begin{itemize}
    \item
        We adjust the concept of forbidden intervals~\cite{BitvectorsMCSAT} to track viable values in \polysat{} (Section~\ref{sec:viable}); 
    \item
        We extract bit-vectors intervals from polynomial (non-linear) inequalities (Section~\ref{sec:intervals}); 
    \item
        We introduce lemmas on-demand for detecting and resolving non-linear conflicts in \polysat{} (Section~\ref{sec:nonlinear}).
    \item We implement \polysat{} directly in the SMT solver Z3~\cite{Z3} and evaluate our work on challenging examples (Section~\ref{sec:experiments}). 
\end{itemize}

\noindent\textit{Paper outline.}
We discuss required preliminaries in Section~\ref{sec:preliminaries} and provide an overview of \polysat{} in Section~\ref{sec:polysat}. We 
describe our main methodological contributions in Sections~\ref{sec:viable}--\ref{sec:nonlinear}
and present our experimental evaluation in Section~\ref{sec:experiments}. 
Section~\ref{sec:conclusion} 
concludes our work. 

\section{Preliminaries\label{sec:preliminaries}}
For a given number of bits~$w > 0$,
we consider bit-vectors of size~$w$ as elements of the ring~$\Zmod{2^w}$ (algebraic representation),
or equivalently as strings of length~$w$ over $\{ 0, 1 \}$ (binary representation).
Throughout the paper,
we write~$w$ for the size of related bit-vectors, when it is clear from the context.
In other cases, we denote the size of~$x$ by~$\bvsize{x}$ explicitly.

For conversion from bit-vectors to integers,
unless explicitly stated otherwise,
we default to the \emph{unsigned} interpretation of bit-vectors,
i.e.,
choose the representatives $\{ 0, 1, \dots, 2^w - 1 \}$ for elements of~$\Zmod{2^w}$.
Negative constants such as $-1$ stand for their equivalent $2^w - 1$.

We write $x \bvule y$ for unsigned comparison of bit-vectors,
and use $x \bvsle y$ to denote signed comparison.
For simplicity of notation, we use ``$=$'' for both object-level equality and meta-level equality.

The basic building blocks of $\polysat{}$ constraints are  \emph{polynomials}, i.e., multiplications and additions of
bit-vector variables and constants.
We emphasize bit-vector multiplication by writing $\cdot$ explicitly.

We write $\bit{x}{i}$ for the $i$-th bit of the bit-vector $x$,
where $\bit{x}{0}$ denotes the least significant bit of $x$.
Let $x \concat y$ denote the concatenation of $x$ and $y$. We write 
 $\extract{x}{h}{l}$, with $0 \leq l \leq h < w$, for 
the \emph{sub-slice} ranging from bit~$h$ to bit~$l$ inclusively, 
i.e., $\extract{x}{h}{l} = \bit{x}{h} \concat \bit{x}{h-1} \concat \dots \concat \bit{x}{l}$.
We call the sub-slices~$\extract{x}{i}{0}$ of~$x$ the \emph{prefixes} of~$x$.

We use half-open \emph{wrapping} intervals over the domain $\Zmod{2^w}$. That is,  for~$l > h$ we define~$\interval{l}{h} \coloneqq \interval{0}{h} \union \interval{l}{2^w}$.
Then, $t \in \interval{l}{h}$ is equivalent to the bit-vector inequality $t - l \bvult h - l$.

\section{\polysat{} in a Nutshell\label{sec:polysat}}

\begin{figure*}[ht!]
\begin{minipage}[b]{0.35\textwidth}
\resizebox{\columnwidth}{!}{
    \begin{tikzpicture}[node distance = 2cm]
        \tikzstyle{component} = [
            rectangle,
            rounded corners,
            minimum width = 1cm,
            minimum height = 1cm,
            text centered,
            draw = black,
        ]
        \tikzstyle{module} = [
            rectangle,
            draw = black,
            dashed,
        ]
        \tikzstyle{arrow} = [
            thick,
            ->,
            > = stealth,
        ]
        \node (boundary) [
            minimum width = 8.75cm, minimum height = 6.5cm, anchor = south west] at (-0.75cm, 1.3) {};

        \node (z3) [module, minimum width = 8cm, minimum height = 8.25cm, anchor = south west] at (0, 1.3) {};
        \node[anchor = south west] at (z3.south west) {\zthree{}};

        \node (egraph) [module, minimum width = 6cm, minimum height = 1.5cm, anchor = south west] at (1.75, 7.75) {};
        \node[anchor = north east, align = right] at (egraph.north east) {\polysat{}\\e-graph~plugin};

        \node (polysat) [module, minimum width = 6cm, minimum height = 5cm,    anchor = south west] at (1.75, 2.5) {};
        \node[anchor = north east, align = right] at (polysat.north east) {\polysat{}\\theory\\solver};

        \node (others)  [module, minimum width = 6cm, minimum height = 0.75cm, anchor = south west] at (1.75, 1.5) {Other Theory Solvers};


        \node (z3core) [component, minimum height = 7.25cm, anchor = south west] at (0.25, 2) {\begin{turn}{90}Z3 Core\end{turn}};

        \begin{scope}[transform canvas = { yshift = 1cm }]
            \draw[arrow] ([xshift = -1cm]z3core.west) -- node[anchor = south, align = center, shift={(-0.5em,0)}] {Input} (z3core.west);
        \end{scope}
        \begin{scope}[transform canvas = { yshift = -1cm }]
            \draw[arrow] (z3core.west) -- node[anchor = north, align = center, shift={(-1em,0)}] {\sat or\\\unsat} ([xshift = -1cm]z3core.west);
        \end{scope}

        \node (slicing) [component, anchor = north west, align = left] at ([xshift = 0.25cm, yshift = -0.25cm]egraph.north west) {
                $\bullet$~Bit-vector slicing\\
                $\bullet$~Fixed values
            };

        \node (search) [component, anchor = north west, align = left] at ([xshift = 0.25cm, yshift = -0.25cm]polysat.north west) {
                Search\\
                $\bullet$~Trail $\Gamma$\\
                $\bullet$~Bit-vector constraints
            };
        \node (viable) [component, anchor = north west, align = left] at ([yshift = -0.2cm]search.south west) {
                Viable Values\\
                $\bullet$~Set of intervals per variable~$x$
            };
        \node (conflict) [component, anchor = north west, align = left] at ([yshift = -0.2cm]viable.south west) {
                Conflict Resolution\\
                $\bullet$~Saturation \\
                $\bullet$~Incremental Linearization\\
                $\bullet$~Bit-Blasting
            };
        \begin{scope}[transform canvas = { xshift = +1.375cm }]
            \draw [arrow] (egraph.south-|viable.north) -- (viable.north);
        \end{scope}
        \begin{scope}[transform canvas = { yshift = +0.125cm }]
            \draw [arrow] (z3core.east|-slicing.west) -- (slicing.west);
        \end{scope}
        \begin{scope}[transform canvas = { yshift = -0.125cm }]
            \draw [arrow] (slicing.west) -- (z3core.east|-slicing.west);
        \end{scope}
        \begin{scope}[transform canvas = { yshift = +0.125cm }]
            \draw [arrow] (z3core.east|-search.west) -- (search.west);
        \end{scope}
        \begin{scope}[transform canvas = { yshift = -0.125cm }]
            \draw [arrow] (search.west) -- (z3core.east|-search.west);
        \end{scope}
        \draw [arrow] ([yshift = 0.5cm]z3core.south east) -- ([xshift = 0.25cm, yshift = -0.25cm]others.north west);
        \begin{scope}[transform canvas = { yshift = -0.25cm }]
            \draw [arrow] ([xshift = 0.25cm, yshift = -0.25cm]others.north west) -- ([yshift = 0.5cm]z3core.south east);
        \end{scope}
    \end{tikzpicture}}
    \caption{%
        \label{fig:overview}%
        \polysat{} Integration%
    }
\end{minipage}
\hfill
\begin{minipage}[b]{0.6\textwidth}
    \small
    \begin{tabular}{llll}
        $p \bvule q$            & unsigned inequality       & \quad
        $\overmul{p}{q}$        & multiplicative overflow   \\
        $x = \band{p}{q}$       & bit-wise \emph{and}       &\quad
        $x = \bshl{p}{q}$       & left shift                \\
        $x = \bor{p}{q}$        & bit-wise \emph{or}        &\quad
        $x = \blshr{p}{q}$      & logical right shift       \\
        $x = \budiv{p}{q}$      & unsigned division         &\quad
        $x = \bashr{p}{q}$      & arithmetic right shift    \\
        $x = \burem{p}{q}$      & unsigned remainder        \\
    \end{tabular}
    \caption{%
        \label{fig:primitive-constraints}%
        Primitive Constraints%
    }

    \vspace{5ex}

    \small
    \begin{tabular}{l@{$~~\leadsto~~$}ll@{$~~\leadsto~~$}l}
        $p \bvult q$        & $\lnot(q \bvule p)$ &\quad
        $p = q$             & $p - q \bvule 0$ \\
        $p \bvsle q$        & $p + 2^{w - 1} \bvule q + 2^{w - 1}$ &\quad
        $p - q$             & $p + (2^{w} - 1) q$ \\
        $\bnot{p}$              & $-p - 1$ &\quad
        $\overadd{p}{q}$    & $p + q \bvult p$ \\
        $\bit{p}{i}$            & $2^{w - 1} \bvule 2^{w - i - 1} p$ \\
    \end{tabular}
    \caption{%
        \label{fig:derived-constraints}%
        Derived Constraints ($w = \bvsize{p} = \bvsize{q}$)%
    }
\end{minipage}
\end{figure*}

\polysat{} serves as a decision procedure for bit-vector constraints and is developed as a theory solver within the SMT solver Z3~\cite{Z3}.
An overview of \polysat{} architecture is given in Figure~\ref{fig:overview},
with further details on key ingredients in Sections~\ref{sec:viable}--\ref{sec:nonlinear}.

In a nutshell, \polysat{} consists of two inter-connected components that interact for theory solving in an SMT setting:
\begin{enumerate}
    \item
        A \emph{bit-vector plugin to the equality graph, in short e-graph~\cite{simplify,egg}}. This plugin 
        handles structural constraints
        that involve multiple bit-widths (concatenation, extraction)
        and determines canonical sub-slices of bit-vectors.
        The \polysat{} e-graph plugin also propagates assigned values across bit-vector slices.
    \item
        A \emph{theory solver}, which handles the remaining constraints by translating
        them into \emph{polynomial constraints}
        (Figure~\ref{fig:primitive-constraints})
        and builds on information from the e-graph plugin
        to search for a satisfiable assignment (Sections~\ref{sec:viable}--\ref{sec:nonlinear}).
\end{enumerate}

From its e-graph, \polysat{} receives Boolean assignments to bit-vector constraints,
and equality propagations between bit-vector terms.
In return, the theory solver of \polysat{} produces a satisfying assignment,
or a conflicting subset of the received constraints. We next discuss these two components,
and then focus on the theory solving aspects of \polysat{} in Sections~\ref{sec:viable}--\ref{sec:nonlinear}.

\subsection{E-graph Plugin}

In SMT solving, an e-graph~\cite{simplify,egg} is typically shared between theory solvers.
The primary purpose of the e-graph is to infer equalities
that follow from congruence reasoning.
For \polysat{}, the e-graph is extended with theory reasoning for bit-vectors.
Theory reasoning is dispatched when the e-graph merges two terms of bit-vector sort.
\polysat{} performs constant propagation over bit-vector extraction and concatenation.
Furthermore, the \polysat{} e-graph establishes equalities between bit-vector ranges.
For example, it infers that $\extract{x}{5}{4} = \extract{x}{1}{0}$
from the equation $\extract{x}{5}{2} = \extract{x}{3}{0}$.

We note that congruence reasoning for bit-vectors was also considered in~\cite{DBLP:conf/fmcad/MollerR98,DBLP:conf/tacas/BjornerP98,DBLP:conf/iccad/BruttomessoS09}. Moreover, 
e-graphs are also used for constant propagation in~\cite{BitvectorsMCSAT}.
The \polysat{} integration of theory plugins to the e-graph structure is generic and not specific to bit-vectors.

\subsection{Theory Solver}

The \emph{propositional search} is driven by the CDCL(T) core
of the SMT solver~\cite{CDCL2,CDCL1}.
\polysat{} receives Boolean assignments to bit-vector constraints
and equality propagations between bit-vector terms.
Both of them are translated into primitive constraints
(cf.~Figure~\ref{fig:primitive-constraints})
and tracked by the \emph{trail}~$\Gamma$.
\polysat{} maintains the invariant that each element of~$\Gamma$ is justified by previous elements,
and that each constraint and variable is assigned at most once in~$\Gamma$.

 \emph{Value search} in \polysat{} assigns viable values (see Section~\ref{sec:viable})
to bit-vector variables,
which are communicated back to the SMT solver core as variable assignment constraints.

\paragraph*{Constraints}
Figures~\ref{fig:primitive-constraints}--\ref{fig:derived-constraints}
list the constraints  that are currently supported in the \polysat{} theory solver,
where~$p$,~$q$ are bit-vector polynomials,
$x$ is a bit-vector variable,
and~$n$ is a bit-vector constant. 
Figure~\ref{fig:primitive-constraints} depicts the \emph{primitive constraints}.
More \emph{expressive constraints} are internally reduced to primitive constraints, see Figure~\ref{fig:derived-constraints}.

\polysat{} uses rewriting to simplify different syntactic forms of equivalent constraints.
In particular, we normalize several forms of equations that may appear in modular arithmetic.
For instance, the constraints $p \bvule 0$, $p \bvult 1$, and $2^w - 1 \bvule p - 1$,
are all normalized to $p = 0$.

Some operations are axiomatized upfront.
For example, to internalize the (unsigned) division $\budiv{x}{y}$,
\polysat{} introduces fresh variables $q \coloneqq \budiv{x}{y}$ and $r \coloneqq \burem{x}{y}$ for the quotient and remainder, respectively.
The main axiom is $x = q y + r$,
but for correctness in bit-vector logic, four more axioms are required:
\begin{align*}
    & \lnot\overmul{q}{y}   && y \neq 0 \limpl r \bvult y \\
    & \lnot\overadd{q y}{r} && y = 0 \limpl q = -1
\end{align*}
where $\lnot\overadd{q y}{r}$ means that the addition $q y + r$ does not overflow,
which can be implemented, e.g., as the constraint $q y \bvule -r-1$.

Constraints of the form~$x = n$, where~$x$ is a variable and~$n$ is a bit-vector constant,
are called \emph{variable assignments}.
Bit-vector terms~$p$ and constraints~$c$ can be evaluated w.r.t.\ the current trail~$\Gamma$,
that is, we substitute the variable assignments in~$\Gamma$ into~$p$ and~$c$, respectively,
and simplify.
As a shorthand, we write~$\eval{p}$ for the evaluation of~$p$ under the current trail.

\paragraph*{Constraint Solving}
The \polysat{} theory solver uses a waterfall model of refinements
to generate lemmas on demand, using the following steps: 

\begin{enumerate}
    \item \label{enum:propagation}
        \emph{Propagation}:
        Value propagation is triggered when a variable is assigned a value
        (Section~\ref{sec:value-propagation}).
    \item \label{enum:viable-conflict}
        \emph{Viable Interval Conflict}:
        If propagation tightens the feasible intervals of a variable to the
        empty set, the solver yields an interval conflict
        (Section~\ref{sec:viable-conflict}).
    \item \label{enum:viable-case-split}
        \emph{Case Split on Viable Candidates}:
        If no further propagation is possible, and there are no interval
        conflicts, the solver picks a value for the next unassigned variable,
        if any.
        It produces a literal $x = n$ for the CDCL solver to case split on,
        with a preference to the phase $x = n$ over $x \neq n$.
        The constant $n$ is chosen to be outside the ranges
        of infeasible intervals stored for $x$ so far
        (Section~\ref{sec:viable-query}).
    \item \label{enum:saturation}
        \emph{Saturation Lemmas}:
        Saturation lemmas let us propagate consequences
        from non-linear constraints
        (Section~\ref{sec:saturation}).
    \item \label{enum:incremental-linearization}
        \emph{Incremental Linearization}:
        Our solver includes incremental linearization rules
        for the cases where variables are $0$, $1$, $-1$, or powers of two
        (Section~\ref{sec:incremental-linearization}).
    \item \label{enum:bit-blasting}
        \emph{Bit-blasting}:
        As a final resort, \polysat{} admits bit-blasting rules
        (Section~\ref{sec:bit-blasting}).
\end{enumerate}

The first three steps above
(steps~\ref{enum:propagation}, \ref{enum:viable-conflict}, \ref{enum:viable-case-split})
operate on linear constraints,
or rather,
a \emph{linear abstraction} of the original constraints,
where non-linear monomials are treated as variables themselves.
If no conflicts arise from the linear abstraction,
then any conflicting non-linear constraints
are handled by the latter stages (steps~\ref{enum:saturation}, \ref{enum:incremental-linearization}, \ref{enum:bit-blasting} above).

A conflict at any stage will cause \polysat{} to return a conflict lemma
to the SMT solver core, which will then backtrack and continue with search.
When control is passed to \polysat{} the next time, theory solving in \polysat{} will begin again in the above step~\ref{enum:propagation} of constraint solving.

\section{Tracking Viable Values\label{sec:viable}}

In the sequel, we discuss the key ingredients of the theory solving component of \polysat{}. A crucial part of the  \polysat{} theory solver tracks for each bit-vector variable~$x$
an over-approximation of the set of feasible values under the current trail~$\Gamma$,
which we call the \emph{viable} values of~$x$.
Specifically,
the set of viable values
is represented as a set of \emph{forbidden intervals},
each of which excludes a certain range of values of~$x$,
and is justified by constraints in the current trail~$\Gamma$.

In \polysat{}, we adapt forbidden intervals from~\cite{BitvectorsMCSAT} and use intervals for propagating and querying  viable values of variables (Sections~\ref{sec:value-propagation}--\ref{sec:viable-query}), and resolving respective conflicts (Section~\ref{sec:viable-conflict}). 
Our approach  extends~\cite{BitvectorsMCSAT} by  computing intervals when the coefficient of~$x$ is not a power of two~(Section~\ref{sec:fi-eq}),
or when the coefficients are different on both sides of an inequality~(Section~\ref{sec:fi-diseq}).

\subsection{Value Propagation\label{sec:value-propagation}}

\polysat{} extracts forbidden intervals from inequalities and overflow
constraints~$c$ that are linear in~$x$ under the current trail~$\Gamma$.
Formally, we determine
an interval~$\interval{l}{u}$ and
side conditions $c_1,\dots,c_n$ that hold under~$\Gamma$
such that
\[
    c \land c_1 \land \dots \land c_n
    \implies
    x \not \in \interval{l}{u}
    .
\]
Intervals are ordered by their starting points,
and we drop intervals that are fully contained in other intervals.
Section~\ref{sec:intervals}
explains how intervals are obtained from constraints.

Value propagation in \polysat{} is triggered when a variable is assigned a value,
or in other words, the solver is presented with a literal $x = n$, where $n$ is a value. 
Propagation is limited to linear occurrences of variables.
For example, if $x$ is assigned $2$, then from $x + y \bvuge 10$,
the non viable intervals for $y$ are updated to $y \not\in \interval{-2}{8}$.
On the other hand, for $xz + y \bvuge 10$, where $x$ occurs in a non-linear term, there is no propagation.
Non-linear propagation in \polysat{} is currently side-stepped because we noticed that it produced very weak lemmas from viable interval conflicts.
Non-linear conflicts are therefore  handled separately, see Section~\ref{sec:nonlinear}.

\subsection{Viable Value Query\label{sec:viable-query}}

To find a viable value for variable~$x$,
we collect the forbidden intervals~$\fiset$ over the prefixes~$\extract{x}{k}{0}$ of~$x$ for $0 \leq k < w$.
In this context, iff an interval $I \in \fiset$ is an interval for $\extract{x}{k}{0}$,
we say \emph{$I$ has bit-width~$k+1$}.
In addition, we consider intervals for variables
that are equivalent to a prefix of~$x$,
as determined by the current state of the e-graph.

In addition to forbidden intervals, we keep track of the set~$C$ of constraints that
are linear in~$x$.
We then invoke Algorithm~\ref{alg:fiLoop}
to either find a value for~$x$ or detect a conflict. To this end, we adjust~\cite{BitvectorsMCSAT}, as follows. 

\begin{algorithm}[!tb]
    \small
    \DontPrintSemicolon
    \caption{\polysat{} Viable Value Query}
    \label{alg:fiLoop}
    \Input{%
        Set of forbidden intervals~$\fiset$,
        set~$C$ of constraints
    }
    \Output{Viable value $x_0$, or a conflict}
    $x_0 \leftarrow x_\mathit{prev}$
    \Comment*{Start at previous viable value}
    $\fijust \leftarrow \emptyseq$
    \Comment*{Justification (sequence of visited intervals)}
    \Loop{%
        \While{$\exists I \in \fiset \text{ such that $x_0 \in I$}$}{%
            Choose such an $I \in \fiset$ with smallest bit-width\label{line:chooseInterval}\;
            $\fijust \leftarrow \pushseq{\fijust}{I}$\label{line:appendJustification}\;
            $x_0 \leftarrow \fnForward(x_0, I)$\label{line:forward}\;
            \lIf{$\fnIsConflict(\fijust)$\label{line:isConflict}}{%
                \Return Conflict~$\fijust$
            }
        }
        \lIf{%
            $x_0 \text{ does not violate any $c \in C$}$%
            \label{line:hasInterval}%
        }{%
            \Return $x_0$%
        }%
        $\fiset \leftarrow \fiset \union \{ \fnComputeInterval(C, x_0) \}$%
        \label{line:computeInterval}\;%
    }%
\end{algorithm}

Algorithm~\ref{alg:fiLoop} starts out with the previous viable value~$x_\mathit{prev}$ of~$x$,
initially set to~$0$.
Then, in the loop of Algorithm~\ref{alg:fiLoop},
we check whether any of the known intervals~$\fiset$
contain the current candidate value~$x_0$ of~$x$.
If that is not the case,
then the current value $x_0$ is compatible with the intervals in~$\fiset$.
We additionally test~$x_0$ for admissibility against the set~$C$ of constraints
(line~\ref{line:hasInterval} of Algorithm~\ref{alg:fiLoop}).
If none of these constraints are violated, the candidate value $x_0$ is returned as viable value for $x$.
Otherwise (line~\ref{line:computeInterval} of Algorithm~\ref{alg:fiLoop}),
$\fnComputeInterval(C, x_0)$ extracts a new
interval that covers~$x_0$ (cf.\ Section~\ref{sec:intervals})
and the search for a viable value of $x$ continues.
If, on the other hand, the current value $x_0$ of $x$ is contained in some forbidden interval,
we choose an interval~$I$ of minimal bit-width among these (line~\ref{line:chooseInterval} of Algorithm~\ref{alg:fiLoop})
and record it in the list~$\fijust$ of justifications (line~\ref{line:appendJustification} of Algorithm~\ref{alg:fiLoop}).

The candidate value~$x_0$ of $x$ is updated to~$\fnForward(x_0, I)$,
the first value after~$x_0$ that is not covered by~$I$
(line~\ref{line:forward} of Algorithm~\ref{alg:fiLoop}).
If a conflict is detected (line~\ref{line:isConflict} of Algorithm~\ref{alg:fiLoop}),
the justifications~$\fijust$ are returned for further processing (see Section~\ref{sec:viable-conflict}).

\subsection{Interval Conflict\label{sec:viable-conflict}}

We detect conflicts by examining the list of justifications~$\fijust$
after appending a new interval~$I$ to $\fijust$.
The condition $\fnIsConflict(\fijust)$ in Algorithm~\ref{alg:fiLoop} is true iff the latest interval~$I$
has already been visited previously,
and no interval of larger bit-width has occurred in between.
Let $I_1,\dots,I_{n+1}$ denote this subsequence of intervals,
where $I_1 = I_{n+1} = I$,
and
let $I_i = \interval{l_i}{h_i}$. 
To block the current assignment to $x$,
\polysat{} creates a conflict lemma from~$I_1,\dots,I_{n+1}$
and reports it to its SMT core.
For simplicity, we only explain here the case where all intervals have same bit-width.

The basic idea of the \polysat{} conflict lemma is the same as in~\cite{BitvectorsMCSAT}:
the union of $I_1,\dots,I_n$ covers the full domain~$\Zmod{w}$,
and the intervals have been chosen such that each upper bound~$h_i$
in contained in the next interval~$I_{i+1}$.
In other words, as long as $h_i \in I_{i+1}$ holds, for all $i$,  and the intervals are valid for $x$,
there can be no feasible value for $x$.
Since the constraints $h_i \in I_{i+1}$ do not contain~$x$ itself,
they are useful for formulating a conflict lemma.
Let $C_i$
denote the set consisting of the constraint and side conditions of~$I_i$.
Then, the \polysat{} conflict lemma is
\[
    \Land_{i=1}^{n} C_i \land \Land_{i=1}^{n} h_i \in I_{i+1} \implies \bot
    .
\]

To illustrate the idea of conflict lemma generation in \polysat{}, consider three intervals
$\interval{l_1}{h_1}$,
$\interval{l_2}{h_2}$,
$\interval{l_3}{h_3}$
whose concrete evaluation under the current trail~$\Gamma$
covers the full domain by
forming the following configuration:
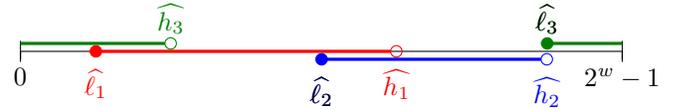
\begin{center}
    \begin{tikzpicture}

        \draw[-] (0,0) -- (8,0);   

        \foreach \x in  {0,8}      
        \draw[shift={(\x,0)},color=black] (0pt,4pt) -- (0pt,-4pt);


        \draw[shift={(0,0)},color=black] (0pt,0pt) -- (0pt,-3pt) node[below] {$0$};
        \draw[shift={(8,0)},color=black] (0pt,0pt) -- (0pt,-3pt) node[below] {$2^w-1$};

        \draw[red,*-o] (0.92,0) -- (5.08,0);
        \draw[red,very thick] (0.92,0) -- (4.92,0);
        \draw[red] (1,-3pt) node[below] {$\widehat{\ell_1}$};
        \draw[red] (5,-3pt) node[below] {$\widehat{h_1}$};
        \draw[blue,shift={(0,-3pt)},*-o] (3.92,0) -- (7.08,0);
        \draw[blue,shift={(0,-3pt)},very thick] (3.92,0) -- (6.92,0);
        \draw[blue,shift={(0,-3pt)}] (4,-3pt) node[below] {$\widehat{\ell_2}$};
        \draw[blue,shift={(0,-3pt)}] (7,-3pt) node[below] {$\widehat{h_2}$};
        \draw[darkgreen,shift={(0,+3pt)},*-] (6.92,0) -- (8,0);
        \draw[darkgreen,shift={(0,+3pt)},very thick] (6.92,0) -- (8,0);
        \draw[darkgreen,shift={(0,+3pt)},-o] (0,0) -- (2.08,0);
        \draw[darkgreen,shift={(0,+3pt)},very thick] (0,0) -- (1.92,0);
        \draw[darkgreen,shift={(0,+3pt)}] (7,0) node[above] {$\widehat{\ell_3}$};
        \draw[darkgreen,shift={(0,+3pt)}] (2,0) node[above] {$\widehat{h_3}$};

        \draw[shift={(0,-3pt)},opacity=0] (4,-3pt) node[below] {$\widehat{\ell_2}$};
        \draw[shift={(0,+3pt)},opacity=0] (7,0) node[above] {$\widehat{\ell_3}$};
    \end{tikzpicture}
\end{center}
Assuming the three intervals are justified by constraints $C_1$, $C_2$, $C_3$, respectively,
the \polysat{} conflict lemma is
\[
    \Land C
    \land h_1 \in \interval{l_2}{h_2}
    \land h_2 \in \interval{l_3}{h_3}
    \land h_3 \in \interval{l_1}{h_1}
    \implies \bot
    ,
\]
where $C \coloneqq C_1 \union C_2 \union C_3$.

\section{Computing Intervals\label{sec:intervals}}

We now describe how forbidden intervals
are extracted from a constraint~$c\in C$
that is linear in the variable~$x$ under consideration.
Intervals may be computed on demand,
relative to a given candidate value (sample point)~$x_0$ of $x$:
the goal is then to find a maximal interval around~$x_0$
of $x$-values that are excluded by~$c$.
In practice, we note the intervals are often not strictly maximal, but as large as reasonably possible to compute.

\subsection{Fixed Bits\label{sec:fi-fixed}}

The e-graph plugin of \polysat{} tracks fixed values for variables and their sub-slices.
If the sample point~$x_0$ contradicts the sub-slice assignment $\extract{x}{h}{l} = n$,
the forbidden interval $\extract{x}{h}{0} \not\in \interval{2^l (n + 1)}{2^l n}$ is created. 
Note that fixed values for sub-slices may also be encoded as inequalities,
for example, as follows:
\begin{center}
    \begin{tabular}{lAl}
        Fixed slice & \multicolumn{2}{c}{Equivalent Constraint}   \\
        \hline\Tstrut
        $\bit{x}{i}$                & 2^{w-i-1} x &\bvuge 2^{w-1}           & \\
        $\extract{x}{h}{0} = n$     & 2^k x &= 2^k n                        & $k \coloneqq w - h - 1$ \\
        $\extract{x}{h}{l} = n$     & 2^k x - 2^{k+l} n &\bvult 2^{k+l}     & $k \coloneqq w - h - 1$ \\
    \end{tabular}
\end{center}
Such (inequality) constraints are turned into appropriate intervals, as described in Section~\ref{sec:fi-eq}. We remark that 
it is not necessary to recover sub-slice assignments by recognizing certain patterns of constraints.

\subsection{Linear Inequality with Equal Coefficients}%
\label{sec:fi-eq}

Given the inequality constraint $p x + q \bvule r x + s$ that is linear in $x$. 
In the cases where either $p$ or $r$ evaluate to $0$ or both to the same value~$a$, the inequality constraint 
is equivalent to an interval constraint~\cite{BitvectorsMCSAT},
according to the following table,
and subject to side conditions $p = \eval{p}$ and $r = \eval{r}$:

\begin{center}
    \begin{tabular}{lll}
        Constraint under $\Gamma$             & Forbidden Interval                & \text{Condition}  \\
        \hline\Tstrut
        $ax + \eval{q} \bvule      \eval{s}$  & $ax \not\in\interval{s-q+1}{-q}$  & $s \neq -1$       \\
        $     \eval{q} \bvule ax + \eval{s}$  & $ax \not\in\interval{-s}{q-s}$    & $q \neq  0$       \\
        $ax + \eval{q} \bvule ax + \eval{s}$  & $ax \not\in\interval{-s}{-q}$     & $q \neq  s$       \\
    \end{tabular}
\end{center}
Assume we have $a x \in \interval{l}{h}$.
Yet, we want to extract an interval on~$x$, rather than on $ax$.
\paragraph*{Case $\mathbf{a = \pm 1}$}
The case $a=1$ trivially leads to such an interval.
In the case $a = -1$ (i.e., $2^w - 1$),
the transformation
\( -x \in \interval{l}{h} \Leftrightarrow x \in \interval{1-h}{1-l} \)
is applied.

\paragraph*{Case $\mathbf{a = \alpha 2^k}$ (reducing the bit-width)}
Consider the case where~$a$ is divisible by~$2^k$ for some~$k > 0$.
Due to the factor~$2^k$,
the upper~$k$ bits of~$x$ do not influence the value of the constraint.
In this case, we consider an interval for the prefix $\extract{x}{w-k-1}{0}$ of $x$:
\[
    \alpha 2^k x \not\in \interval{l}{h}
    \iff
    \begin{cases}
        \alpha \extract{x}{w-k-1}{0} \not\in \interval{l'}{h'}  & \text{if $l' \neq h'$} \\
        0 \not\in \interval{l}{h}    & \text{otherwise}
    \end{cases}
\]
where
\(
    \beta' \coloneqq
    \roundup{\frac{\beta}{2^k}} \modop 2^{w-k}
\)
for $\beta \in \{ l, h \}$.

\paragraph*{Other values of $\mathbf{a}$}
For other values of $a$, in general, multiple disjoint intervals exist.
We extract intervals around a sample point~$x_0$ on demand,
i.e., given concrete values $a, x_0, l, h \in \Zmod{2^w}$
such that $a x_0 \in \interval{l}{h}$,
the task is to compute the maximal $x$-interval~$\interval{x_l}{x_h}$
such that $a x \in \interval{l}{h}$ for all $x \in \interval{x_l}{x_h}$.
To compute~$x_l$ and~$x_h$, we move the problem into the integers~$\mathbb{Z}$
and work with non-wrapping intervals.
Operations until the end of this subsection
are therefore to be understood as operations in~$\mathbb{Z}$.

\tikzmath{
    real \M, \iM;  
    function project(\x) {
        return \x * \iM / \M;
    };
    \iM = 6;
    \M = 1024;
    \a = 197;
    \x = 636;
    \k = 14;
    \lo = 50;
    \hi = 1023;  
    \ia = project(\a);
    \iax = project(364);
    \ilo = project(\lo);
    \ihi = project(\hi);
    \dotr = 0.08 cm;
}
Let $w$ be a fixed bit-width and let $m \coloneqq 2^w$.
Assume values $a, x_0, l, h \in \mathbb{Z}$
are given such that
$1 \leq a < m$,
$-m < l \leq h < m$,
and
$a x_0 \modop m \in \intervalcc{l}{h}$.
Furthermore, the length of the interval should be less than $m$, i.e., $h - l + 1 < m$
(otherwise the computation is unnecessary because the corresponding modular interval covers the whole domain).
The goal is to find the minimal~$x_l$ and the maximal~$x_h$
such that $a x \modop m \in \intervalcc{l}{h}$
for all $x \in \intervalcc{x_l}{x_h}$.

Let $k_0 \in \mathbb{Z}$ such that $l \leq a x_0 + k_0 m \leq h$.
\newcommand{\norm}[1]{\langle #1 \rangle}
To simplify notation, define $\norm{x} \coloneqq x + k_0 m$.
The initial configuration is illustrated by the following diagram:
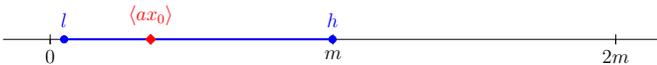
\begin{center}
\resizebox{\columnwidth}{!}{
    \begin{tikzpicture}
        \draw[-] (-1,0) -- (13,0);   

        \foreach \x in  {0,6,12}     
        \draw[shift={(\x,0)},color=black] (0pt,3pt) -- (0pt,-3pt);

        \draw[shift={(0,0)},color=black]  (0pt,0pt) -- (0pt,-3pt) node[below,fontscale=1] {$0$};
        \draw[shift={(\iM,0)},color=black]  (0pt,0pt) -- (0pt,-3pt) node[below,fontscale=1] {$m$};
        \draw[shift={(2*\iM,0)},color=black] (0pt,0pt) -- (0pt,-3pt) node[below,fontscale=1] {$2m$};

        \draw[blue,*-*]         (\ilo cm - \dotr,0) -- (\ihi cm + \dotr,0);
        \draw[blue,very thick]  (\ilo cm + \dotr,0) -- (\ihi cm - \dotr,0);
        \draw[blue]             (\ilo cm, +4pt)     node[above,fontscale=1] {$l$};
        \draw[blue]             (\ihi cm, +4pt)     node[above,fontscale=1] {$h$};

        \node at (\iax,0) [diamond,fill,red,inner sep=1.75pt] {};
        \draw[red]   (\iax,5pt)    node[above,fontscale=1] {$\norm{a x_0}$};
    \end{tikzpicture}}
\end{center}

Since we are ultimately interested in the modular interval~$\intervalcc{l}{h} \modop m$ over $\Zmod{m}$,
we consider the set of all representatives of elements of that interval,
i.e., the union of $\intervalcc{l}{h} + im$ for all $i\in\mathbb{Z}$,
as depicted in the following diagram.
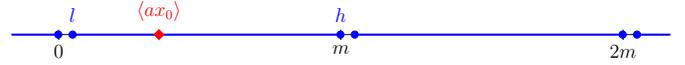
\begin{center}
\resizebox{\columnwidth}{!}{
    \begin{tikzpicture}
        \draw[-] (-1,0) -- (13,0);   

        \foreach \x in  {0,6,12}     
        \draw[shift={(\x,0)},color=black] (0pt,3pt) -- (0pt,-3pt);

        \draw[shift={(0,0)},color=black]  (0pt,0pt) -- (0pt,-3pt) node[below,fontscale=1] {$0$};
        \draw[shift={(\iM,0)},color=black]  (0pt,0pt) -- (0pt,-3pt) node[below,fontscale=1] {$m$};
        \draw[shift={(2*\iM,0)},color=black] (0pt,0pt) -- (0pt,-3pt) node[below,fontscale=1] {$2m$};

        \draw[shift={(\iM,0)},blue,*-*]         (\ilo cm - \dotr,0) -- (\ihi cm + \dotr,0);
        \draw[shift={(\iM,0)},blue,very thick]  (\ilo cm + \dotr,0) -- (\ihi cm - \dotr,0);
        \draw[shift={(2*\iM,0)},blue,*-]          (\ilo cm - \dotr,0) -- (1,0);
        \draw[shift={(2*\iM,0)},blue,very thick]  (\ilo cm + \dotr,0) -- (1,0);
        \draw[shift={(-\iM,0)},blue,-*]          (5,0) -- (\ihi cm + \dotr,0);
        \draw[shift={(-\iM,0)},blue,very thick]  (5,0) -- (\ihi cm - \dotr,0);

        \draw[blue,*-*]         (\ilo cm - \dotr,0) -- (\ihi cm + \dotr,0);
        \draw[blue,very thick]  (\ilo cm + \dotr,0) -- (\ihi cm - \dotr,0);
        \draw[blue]             (\ilo cm, +4pt)     node[above,fontscale=1] {$l$};
        \draw[blue]             (\ihi cm, +4pt)     node[above,fontscale=1] {$h$};

        \node at (\iax,0) [diamond,fill,red,inner sep=1.75pt] {};
        \draw[red]   (\iax,5pt)    node[above,fontscale=1] {$\norm{a x_0}$};
    \end{tikzpicture}}
\end{center}

The underlying idea of our procedure is to look
at each interval representative $\intervalcc{l}{h} + im$ separately
(intuitively, as a region where no overflow occurs)
and take advantage of periodicity after each overflow.

In the first step, we compute the minimal $x_l'$ and the maximal $x_h'$
such that $l \leq \norm{a x} \leq h$ for all $x \in \intervalcc{x_l'}{x_h'}$.
Intuitively, $\intervalcc{x_l'}{x_h'}$ is the maximal $x$-interval around $x_0$
such that no overflow occurs among the corresponding multiples of~$a$.
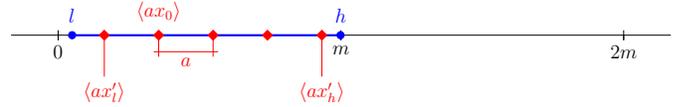
\begin{center}
\resizebox{\columnwidth}{!}{
    \begin{tikzpicture}
        \draw[-] (-1,0) -- (13,0);   

        \foreach \x in  {0,6,12}      
        \draw[shift={(\x,0)},color=black] (0pt,3pt) -- (0pt,-3pt);

        \draw[shift={(0,0)},color=black]  (0pt,0pt) -- (0pt,-3pt) node[below,fontscale=1] {$0$};
        \draw[shift={(\iM,0)},color=black]  (0pt,0pt) -- (0pt,-3pt) node[below,fontscale=1] {$m$};
        \draw[shift={(2*\iM,0)},color=black] (0pt,0pt) -- (0pt,-3pt) node[below,fontscale=1] {$2m$};

        \draw[blue,*-*]         (\ilo cm - \dotr,0) -- (\ihi cm + \dotr,0);
        \draw[blue,very thick]  (\ilo cm + \dotr,0) -- (\ihi cm - \dotr,0);
        \draw[blue]             (\ilo cm, +4pt)     node[above,fontscale=1] {$l$};
        \draw[blue]             (\ihi cm, +4pt)     node[above,fontscale=1] {$h$};

        \foreach \i in  {-1,0,1,2,3}
        \node[shift={(\i*\ia,0)}] at (\iax,0) [diamond,fill,red,inner sep=1.75pt] {};

        \draw[red]   (\iax      ,5pt)      node[above,fontscale=1] {$\norm{a x_0 }$};
        \draw[red]   (\iax-\ia  ,-25pt)    node[below,fontscale=1] {$\norm{a x_l'}$};
        \draw[red]   (\iax+3*\ia,-25pt)    node[below,fontscale=1] {$\norm{a x_h'}$};

        \draw[red,-] (\iax-\ia  ,0) -- (\iax-\ia    ,-25pt);
        \draw[red,-] (\iax+3*\ia,0) -- (\iax+3*\ia  ,-25pt);

        \draw[red,-] (\iax cm - 3pt, -10pt) -- (\iax cm + \ia cm + 3pt, -10pt) node[midway,below,fontscale=1] {$a$};
        \draw[red,-] (\iax    ,0) -- (\iax    ,-13pt);
        \draw[red,-] (\iax+\ia,0) -- (\iax+\ia,-13pt);
    \end{tikzpicture}}
\end{center}

However, the interval $\intervalcc{x_l'}{x_h'}$ is often far from optimal,
causing repeated queries over the same constraint in Alg.~\ref{alg:fiLoop}.
In case of the upper bound, this means that $\norm{a (x_h' + 1)}$
is contained in the next interval representative $\intervalcc{l}{h}+m$.
The following diagram illustrates the multiples of~$a$ across several interval representatives.
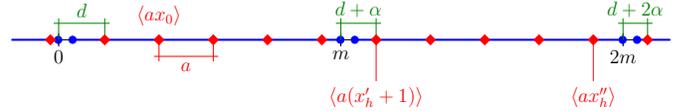
\begin{center}
\resizebox{\columnwidth}{!}{
    \begin{tikzpicture}
        \draw[-] (-1,0) -- (13,0);   

        \foreach \i in  {0,1,2}      
        \draw[shift={(\i*\iM,0)},color=black] (0pt,4pt) -- (0pt,-4pt);

        \draw[shift={(0,0)},color=black]  (0pt,0pt) -- (0pt,-3pt) node[below,fontscale=1] {$0$};
        \draw[shift={(\iM,0)},color=black]  (0pt,0pt) -- (0pt,-3pt) node[below,fontscale=1] {$m$};
        \draw[shift={(2*\iM,0)},color=black] (0pt,0pt) -- (0pt,-3pt) node[below,fontscale=1] {$2m$};

        \draw[blue,*-*]         (\ilo cm - \dotr,0) -- (\ihi cm + \dotr,0);
        \draw[blue,very thick]  (\ilo cm + \dotr,0) -- (\ihi cm - \dotr,0);
        \draw[shift={(\iM,0)},blue,*-*]         (\ilo cm - \dotr,0) -- (\ihi cm + \dotr,0);
        \draw[shift={(\iM,0)},blue,very thick]  (\ilo cm + \dotr,0) -- (\ihi cm - \dotr,0);
        \draw[shift={(2*\iM,0)},blue,*-]          (\ilo cm - \dotr,0) -- (1,0);
        \draw[shift={(2*\iM,0)},blue,very thick]  (\ilo cm + \dotr,0) -- (1,0);
        \draw[shift={(-\iM,0)},blue,-*]          (5,0) -- (\ihi cm + \dotr,0);
        \draw[shift={(-\iM,0)},blue,very thick]  (5,0) -- (\ihi cm - \dotr,0);

        \draw[red]   (\iax      ,5pt)      node[above,fontscale=1] {$\norm{a x_0 }$};
        \draw[red]   (\iax+4*\ia,-25pt)    node[below,fontscale=1] {$\norm{a (x_h'+1)}$};
        \draw[red]   (\iax+8*\ia,-25pt)    node[below,fontscale=1] {$\norm{a x_h''}$};
        \draw[red,-] (\iax+4*\ia,0) -- (\iax+4*\ia  ,-25pt);
        \draw[red,-] (\iax+8*\ia,0) -- (\iax+8*\ia  ,-25pt);

        \draw[darkgreen,-] (        -3pt,   +10pt) -- (\iax cm - \ia cm + 3pt,   +10pt) node[midway,above,fontscale=1] {$d$};
        \draw[darkgreen,-] (0       ,           0) -- (0       ,                 +13pt);
        \draw[darkgreen,-] (\iax-\ia,           0) -- (\iax-\ia,                 +13pt);

        \draw[darkgreen,-] (\iM cm - 3pt,   +10pt) -- (\iax cm + 4*\ia cm + 3pt, +10pt) node[midway,above,fontscale=1] {$d + \alpha$};
        \draw[darkgreen,-] (\iM cm      ,       0) -- (\iM cm      ,             +13pt);
        \draw[darkgreen,-] (\iax + 4*\ia,       0) -- (\iax + 4*\ia,             +13pt);

        \draw[darkgreen,-] (2*\iM cm - 3pt, +10pt) -- (\iax cm + 9*\ia cm + 3pt, +10pt) node[midway,above,fontscale=1] {$d + 2\alpha$};
        \draw[darkgreen,-] (2*\iM cm      ,     0) -- (2*\iM cm                , +13pt);
        \draw[darkgreen,-] (\iax + 9*\ia  ,     0) -- (\iax + 9*\ia            , +13pt);

        \foreach \i in  {-2,-1,0,1,2,3,4,5,6,7,8,9}
        \node[shift={(\i*\ia,0)}] at (\iax,0) [diamond,fill,red,inner sep=1.75pt] {};

        \draw[red,-] (\iax cm - 3pt, -10pt) -- (\iax cm + \ia cm + 3pt, -10pt) node[midway,below,fontscale=1] {$a$};
        \draw[red,-] (\iax    ,0) -- (\iax    ,-13pt);
        \draw[red,-] (\iax+\ia,0) -- (\iax+\ia,-13pt);
    \end{tikzpicture}}
\end{center}

The situation in the second interval~$\intervalcc{l}{h}+m$
is very similar to the initial setting.
However, the multiples of $a$
(depicted by red diamonds)
have shifted by some amount~$\alpha$
relative to the interval.

In the example illustrated in the diagrams we have
$\alpha < 0$, i.e.,
with each overflow, the multiples of $a$
drift to the left (relative to the interval).
With different parameters, $\alpha = 0$ (no drift)
and $\alpha > 0$ (drift to the right) are also possible.

For $\alpha < 0$,
we keep overflowing until the leftmost multiple of $a$ drifts outside the interval.
For $\alpha > 0$, similarly for the rightmost multiple of $a$
(in this case, the final considered interval will be irregular in the sense that it contains one fewer multiple of $a$).

In case $\alpha = 0$,
the situation for each interval representative is exactly the same,
and we conclude no upper bound $x_h$ exists
(which means the final $x$-interval over $\Zmod{m}$ will be the full domain).

We have described our method to compute the upper bound~$x_h$.
The lower bound~$x_l$ can be computed analogously.
In fact, \polysat{} reduces the computation of~$x_l$ to the computation of~$x_h$
by mirroring the initial configuration and the result across~$0$.
Let~$f$ denote the procedure for calculating~$x_h$,
i.e., $x_h = f(x_0, a, l, h, m)$.
Then
\(
    x_l = -f(-x_0, a, -h, -l, m)
\).

Even though this method works well in practice,
some limitations remain.
The interval extension ends as soon as one of
the red diamonds is outside the blue interval.
This is by specification, but it does mean that this
method is only helpful when the gap between blue intervals (i.e., $m - (h - l)$)
is less than the distance between red diamonds (i.e., $a$).

\subsection{Linear Inequality with Different Coefficients}%
\label{sec:fi-diseq}

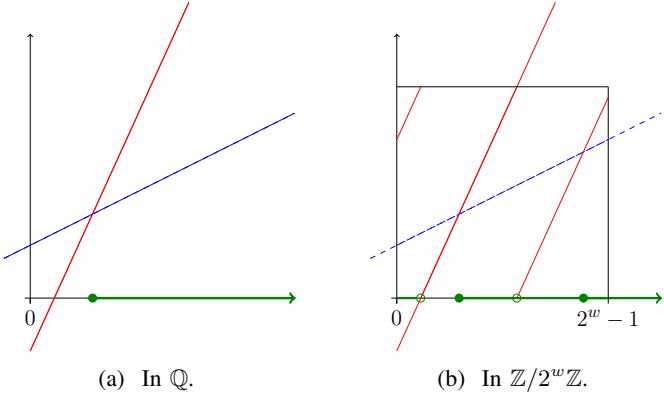
\begin{figure}
\begin{minipage}{\columnwidth}

    \begin{subfigure}[b]{0.45\textwidth}
    \resizebox{\textwidth}{!}{
        \begin{tikzpicture}
            \draw[->,very thin] (-3pt,0) -- (5,0);   
            \draw[->,very thin] (0,-3pt) -- (0,5);   
            \draw[shift={(0,0)},color=black] (0pt,0pt) -- (0pt,-3pt) node[below,fontscale=1.5] {$0$};
            \draw[-,blue,dashed] (-0.5,0.75) -- (5,3.5);
            \draw[-,red] (0,-1.0) -- (3,5.6000000000000005);
            \draw[darkgreen,*-] (1.0964705882352939,0) -- (5,0);
            \draw[darkgreen,very thick,->] (1.176470588235294,0) -- (5.02,0);
            \draw[-,blue,opacity=0.05] (-0.5,0.75) -- (5,3.5);
            \draw[-,red,opacity=0.05] (0,-1.0) -- (3,5.6000000000000005);
            \draw[darkgreen,very thick,->,opacity=0] (1.176470588235294,0) -- (5.02,0);
        \end{tikzpicture}}
        \caption{%
            \label{fig:fi-diseq-rational}
            In $\mathbb{Q}$.
        }
    \end{subfigure}
    \hfill
    \begin{subfigure}[b]{0.45\textwidth}
    \resizebox{\textwidth}{!}{
        \begin{tikzpicture}
            \draw[->,very thin] (-3pt,0) -- (5,0);   
            \draw[->,very thin] (0,-3pt) -- (0,5);   
            \draw[-,blue,opacity=0.05,dashed] (-0.5,0.75) -- (5,3.5);
            \draw[-,red,opacity=0.05] (0,-1.0) -- (3,5.6000000000000005);
            \draw[darkgreen,very thick,->,opacity=0] (1.176470588235294,0) -- (5.02,0);
            \draw[-,very thin] (0,4) -- (4,4);
            \draw[-,very thin] (4,0) -- (4,4);
            \draw[shift={(0,0)},color=black] (0pt,0pt) -- (0pt,-3pt) node[below,fontscale=1.5] {$0$};
            \draw[shift={(4,0)},color=black] (0pt,0pt) -- (0pt,-3pt) node[below,fontscale=1.5] {$2^w-1$};
            \draw[-,blue,dashed] (0,1.0) -- (4,3.0);
            \draw[-,red] (0,3.0) -- (0.45454545454545453,4.0);
            \draw[-,red] (0.45454545454545453,0.0) -- (2.2727272727272725,4.0);
            \draw[-,red] (2.2727272727272725,0.0) -- (4,3.8000000000000007);
            \draw[darkgreen,*-] (1.0964705882352939,0) -- (1.7245989304812832,0);
            \draw[darkgreen,very thick] (1.176470588235294,0) -- (1.7245989304812832,0);
            \draw[darkgreen,-o] (1.7245989304812832,0) -- (2.3527272727272726,0);
            \draw[darkgreen,very thick] (1.7245989304812832,0) -- (2.1927272727272724,0);
            \draw[darkgreen,*-] (3.4494117647058817,0) -- (4,0);
            \draw[darkgreen,very thick] (3.529411764705882,0) -- (4,0);
            \draw[darkgreen,-o] (0,0) -- (0.5345454545454545,0);
            \draw[darkgreen,very thick] (0,0) -- (0.3745454545454545,0);
        \end{tikzpicture}}
        \caption{%
            \label{fig:fi-diseq-modular}
            In $\mathbb{Z}/2^w\mathbb{Z}$.
        }
    \end{subfigure}
    \caption{%
        \label{fig:fi-diseq}
        Example for extracting intervals from an inequality constraint~$px+q \bvule rx+s$ with different variable coefficients.
        The blue dashed line plots $\eval{p}x+\eval{q}$, and
        the red continuous line is $\eval{r}x+\eval{s}$.
    }
\end{minipage}
\end{figure}

Consider an inequality $c$ of the form $px + q \bvule rx + s$
with~$\eval{p} \neq \eval{r}$.
Here, we need to  find the largest $x$-interval
around a sample point $x_0$ where $c$ is satisfied.
As Figure~\ref{fig:fi-diseq-rational} shows for an example,
the corresponding problem is easily solved
over infinite domains, such as  rationals, by computing the intersection point
of the left- and right-hand side of the inequality.
The interval extends from the intersection point towards infinity.

However, in modular arithmetic, the left-hand side and the right-hand side of $c$
do not represent continuous lines; instead, they wrap around at $2^w$
as seen in Figure~\ref{fig:fi-diseq-modular}.
The intervals extend from an intersection point to the next wraparound point.
We compute and return the interval containing $x_0$.

However, \polysat{} computes only the intersection/wraparound points nearest to~$x_0$.
In some configurations, the gap between one interval to the next
(i.e., between the green lines in Figure~\ref{fig:fi-diseq-modular})
does not contain an integer, which means the obtained $x$-interval is not maximal.
This method works best when the coefficients of~$x$ are near $0$ or $2^w$.

Consider an inequality $px+q \bvule rx+s$
with $p,q,r,s \in \Zmod{2^w}$
such that $p \neq 0$, $r \neq 0$ and $p \neq r$.
Let $x_0 \in \Zmod{2^w}$ be a sample value that violates the constraint,
i.e., such that $(p x_0 + q) \modop 2^w > (r x_0 + s) \modop 2^w$
(to avoid confusion, we write ``$\modop$'' operations in this section explicitly).

The goal is to find a maximal $x$-interval around $x_0$
whose elements all violate the constraint, i.e.,
we want to find the minimal~$x_l$ and the maximal $x_h$
such that $x_l \leq x_0 \leq x_h$
and $(p x + q) \modop 2^w > (r x + s) \modop 2^w$
for all $x \in \intervalcc{x_l}{x_h}$.

In the following, we explain our method for extracting such intervals,
however, we cannot yet guarantee to obtain a maximal interval in all cases.
As illustrated in Figure~\ref{fig:fi-diseq},
we extrapolate the left-hand side (LHS) and the right-hand side (RHS) of the constraint
using standard arithmetic until the next overflow point,
and extract the maximal interval that can be obtained without overflow.

Let us define the abbreviations
$a \coloneqq (p x_0 + q) \modop 2^w$
and
$b \coloneqq (r x_0 + s) \modop 2^w$.
From now on, we view $p,q,r,s,a,b$ as values over the rationals~$\mathbb{Q}$
by choosing the representative in the interval~$\interval{0}{2^w}$.

To compute a safe upper bound $x_h = x_0 + \delta_h$,
we find the maximal $\delta_h\in\mathbb{Z}$ satisfying the following conditions:
\begin{itemize}
    \item $\delta_h \geq 0$, i.e., it should be an \emph{upper} bound,
    \item $\forall x . ( 0 \leq x \leq \delta_h \rightarrow 2^w > a + p x > b + r x \geq 0 )$,
        i.e., the LHS and RHS do not overflow within the interval and the constraint is violated for all values,
    \item $x_0 + \delta_h < 2^w$, i.e., the upper bound does not overflow.
\end{itemize}

After several transformations, we obtain the formula
\[
    \delta_h =
    \min \left(
        \Bigl\{ 2^w - x_0, \bigl\lceil \frac{2^w - a}{p} \bigr\rceil \Bigr\}
        \cup
        \Bigl\{ \bigl\lceil \frac{a - b}{r-p} \bigr\rceil \Bigm\vert r > p \Bigr\}
    \right) - 1
    .
\]

Similarly, we obtain a safe lower bound $x_l = x_0 - \delta_l$,
by finding the maximal $\delta_l\in\mathbb{Z}$ such that:
\begin{itemize}
    \item $\delta_l \geq 0$ (it should be a \emph{lower} bound),
    \item $\delta_l \leq x_0$ (lower bound does not overflow),
    \item $\forall x . ( 0 \leq x \leq \delta_l \rightarrow 2^w > a - p x > b - r x \geq 0 )$.
\end{itemize}
A sequence of transformations leads us to the formula
\[
    \delta_l =
    \min \left(
        \Bigl\{ x_0 + 1, \bigl\lceil \frac{b+1}{r} \bigr\rceil \Bigr\}
        \cup
        \Bigl\{ \bigl\lceil \frac{a-b}{p-r} \bigr\rceil \Bigm\vert p > r \Bigr\}
    \right) - 1
    .
\]

At the beginning of this section,
we embedded the coefficients $p,q$ from $\Zmod{2^w}$
into $\mathbb{Q}$ by choosing the representative in the interval~$\interval{0}{2^w}$.
However, whenever $p$ or $q$ is a large value near $2^w$ we may obtain better bounds by
interpreting them as negative numbers, i.e., choose the representative in the interval~$\interval{-2^w}{0}$ instead.
To obtain a uniform formula, we can simply plug in $p-2^w$ and $q-2^w$ for $p$ and $q$ (or just one of them), respectively,
in the formulas above.
In total, this gives us four different ways to estimate each bound.
Since each of these computations finds a safe bound, we choose the best among them.

Finally, if we want to compute such bounds for a strict inequality $px+q \bvult rx+s$,
we only have to change the strictness of one inequality in our initial conditions,
i.e., replace $a \pm px > b \pm rx$ by $a \pm px \geq b \pm rx$.
In the final formulas, this manifests as replacing $a-b$ in the numerator by $a-b+1$;
otherwise, the results are unchanged.

\subsection{Projecting intervals to sub-slices}

Since value assignments are propagated
eagerly across bit-vector slices
by the e-graph component of \polysat{},
in some cases, a bit-vector variable is assigned to a value
that contradicts an interval on a super-slice of the variable.
Such contradictions may also be caused by the e-graph,
because it does not take into account intervals when merging nodes.

Let $x := y \concat z$ s.t. $\bvsize{y} = u$ and $\bvsize{z} = v$.
Given the forbidden interval $x \not\in \interval{l}{h}$,
then $2^v y + z \not \in \interval{l}{h}$.
We can learn intervals for $y$ and $z$ via the following \polysat{} lemmas. 

\begin{lemma}[General Intervals]
    \label{lemma:projected-general-intervals}
    In case no fixed value is known for the other sub-slice,
    it is possible to learn an interval
    as long as $\interval{l}{h}$ is big enough.
    \begin{alignat}{2}
        &\ilen{\interval{l}{h}} \geq 2^u        &&\implies y \not\in \interval{l_y}{h_y} \label{eq:chop-lower-general} \\
        &\ilen{\interval{l}{h}} > 2^{u+v} - 2^v &&\implies z \not\in \interval{l_z}{h_z} \label{eq:chop-upper-general}
    \end{alignat}
    where
    $l_y \coloneqq \roundup{\frac{l}{2^v}} \modop 2^v$,
    $h_y \coloneqq \rounddown{\frac{h}{2^v}}$,
    $l_z \coloneqq l \modop 2^v$, and
    $h_z \coloneqq h \modop 2^v$.
\end{lemma}

\begin{lemma}[Specific Intervals]
    \label{lemma:projected-specific-intervals}
    If the other sub-slice has a fixed value,
    a larger interval can be projected
    \cite[Figure~1]{BitvectorsMCSAT}.
    \begin{alignat}{2}
        &z = n \land l_y \neq h_y                                       &&\implies y \not\in[l_y, h_y[              \label{eq:chop-lower-fixed-n}       \\
        &z = n \land l_y = h_y \land h_y 2^v + n \in \interval{l}{h}    &&\implies \bot                             \label{eq:chop-lower-fixed-full}    \\
        &y = n \land l_z \neq h_z                                       &&\implies z \not\in \interval{l_z}{h_z}    \label{eq:chop-upper-fixed-n}       \\
        &y = n \land l_z = h_z \land n 2^v \in \interval{l}{h}          &&\implies \bot                             \label{eq:chop-upper-fixed-full}
    \end{alignat}
    where ($\beta \in \{l, h\}$)
    \begin{align*}
        \beta_y &\coloneqq
        \bigroundup{\frac{(\beta - n) \modop 2^{u+v}}{2^v}} \modop 2^u,
        \\
        \beta_z &\coloneqq
        \begin{cases}
            \beta \modop 2^v   & \text{if $\rounddown{\frac{\beta}{2^v}} = n$}, \\
            0               & \text{otherwise}.
        \end{cases}
    \end{align*}
\end{lemma}

These projections are applied iteratively in \polysat{} to derive intervals
for arbitrary sub-slices.
At each step, a choice is made between
Lemmas~\ref{lemma:projected-general-intervals}--\ref{lemma:projected-specific-intervals},
depending on whether a fixed value is available at the required decision level.
\begin{example}
    We can use the above to find an interval~$I$ such that
    \(
        x = 0 \concat y \concat z
        \land
        \extract{z}{15}{8} = 123
        \land
        x \not\in \interval{300007}{0}
    \)
    implies~$y \not\in I$,
    where $\bvsize{x} = 64$ and $\bvsize{y} = \bvsize{z} = 16$.
    \begin{itemize}
        \item
            First, apply~\eqref{eq:chop-upper-fixed-n}
            to obtain $y \concat z \not\in \interval{300007}{0}$.
        \item
            Next, with~\eqref{eq:chop-lower-general}
            we obtain $y \concat \extract{z}{15}{8} \not\in \interval{1253}{0}$.
        \item
            Finally, with~\eqref{eq:chop-lower-fixed-n}
            we obtain $y \not\in \interval{5}{0}$.
    \end{itemize}
\end{example}

\section{Non-Linear Conflicts\label{sec:nonlinear}}

Non-linear conflicts are handled in \polysat{} by saturation, incremental linearization, and bit-blasting.
Saturation, incremental linearization and bit-blasting are  postponed until all variables are assigned
values and there are no conflicts detected by propagating bounds on linear constraints.

\subsection{Saturation Lemmas\label{sec:saturation}}

Saturation lemmas  propagate consequences from non-linear constraints.
The consequences are considered ``simpler'',
when they are linear or if they contain fewer variables. 
Saturation lemmas, given in Lemmas~\ref{lemma:saturation-modulo-multiplication}--\ref{lemma:parity-saturation}, are added by \polysat{} if their non-linear constraints are in the assertion trail and they evaluate to false
under the current assignment in~$\Gamma$.

\begin{lemma}[Saturation Modulo Multiplication Inequalities]
\label{lemma:saturation-modulo-multiplication}
We give an excerpt of possible saturation rules.
An extended list can be found in Appendix~\ref{appendix:additional-lemmas}.
\[
\begin{array}{llllllllll}
p x \bvult q x                  &\implies&& \overmul{p}{x}      &\lor& p \bvult q   \\
p x \bvult q x                  &\implies&& \overmul{-q}{x}     &\lor& p \bvult q   \\
p x \bvult q x                  &\implies&& \overmul{q}{-x}     &\lor& p \bvugt q   \\
~                               &&\lor&     p = 0                                   \\
p x \bvult q x                  &\implies&& \overmul{-p}{-x}    &\lor& p \bvugt q   \\
~                               &&\lor&     p = 0                                   \\
p x \bvule q x                  &\implies&& \overmul{p}{x}      &\lor& p \bvule q   \\
~                               &&\lor&     x = 0                                   \\
px + s \bvule q &\implies& & \overmul{p}{x}   &\lor & \overadd{px}{s}  \\
&                       & \lor & p r \bvule q &\lor & x \bvult r       \\
p \bvule x \land qx \bvule r    &\implies&& \overmul{q}{x} & \lor& pq \bvule r  \\
p \bvule x \land qx \bvult r    &\implies&& \overmul{q}{x} & \lor& pq \bvult r  \\
p \bvule qx \land x \bvule r    &\implies&& \overmul{q}{r} & \lor& p \bvule qr  \\
p \bvult qx \land x \bvule r    &\implies&& \overmul{q}{r} & \lor& p \bvult qr  \\
\end{array}
\]
Note that these rules do not require $x \not\in p, q, r, s$,
so they can be applied even when the degree of $x$ is larger than $1$.
\end{lemma}

Next, we can connect overflow constraints with multiplications or decompose them to linear inequalities.
\begin{lemma}[Overflow Saturation]
\label{lemma:saturation-modulo-overflow}
\[
    \begin{array}{lll}
        \lnot\overmul{p}{q} \land q \neq 0          & \implies & p \bvule p \cdot q         \\
        \bar{0}p \cdot \bar{0}q \bvuge 2^w          & \implies & \overmul{p}{q}             \\
        \overmul{p}{q} \land \lnot\overmul{r}{s}    & \implies & p \bvugt r \lor q \bvugt s \\
        \overmul{p}{q} \land p \bvuge q             & \implies & p \bvuge \roundup{\sqrt{2^w}}      \\
        \lnot\overmul{p}{q} \land p \bvuge q        & \implies & q \bvult \rounddown{\sqrt{2^w}}    \\
    \end{array}
\]
where $\bar{0}p$ and $\bar{0}q$
stands for a zero-extension with at least one bit of $p$ and $q$, respectively.
Note that here $w = \bvsize{p} = \bvsize{q} > 1$, since for $w=1$ multiplication overflow is impossible.
\end{lemma}

Variables can in some cases be resolved, producing constraints
that are free of resolved variables.
\begin{lemma}[Saturation Modulo Equalities]
    \label{lemma:saturation-modulo-equalities}
    \[
    \begin{array}{llll}
        ax + b = 0 \land cx + d = 0 & \implies & ad - bc = 0 \\
        ax + b = 0 \land c[x] & \implies & c[-b\cdot a^{-1}] & \mbox{if $a$ is odd}
    \end{array}
    \]
    where~$c[x]$ may be any constraint containing~$x$.
    Note that the multiplicative inverse~$a^{-1}$ of~$a$ in~$\Zmod{2^w}$
    exists if and only if~$a$ is odd.
\end{lemma}

Finally, let us define the \emph{parity} of a bit-vector~$x$
as the largest number~$i \in \{ 0, \dots, w \}$ such that~$2^i$ divides~$x$.
The parity of a bit-vector can be constrained by a linear inequality,
where~$\parity(p) \geq i \Longleftrightarrow  p 2^{w - i} = 0$ for~$0 < i \leq w$.

\begin{lemma}[Parity Saturation]%
    \label{lemma:parity-saturation}%
    Parity inequalities can be used to constrain values of multipliers.%
    \[%
    \begin{array}{lcl}
        p \cdot q = 0 & \implies & \parityOf{p} + \parityOf{q} \geq w \\
        p \cdot q = 1 & \implies & \parityOf{p} = 0 \\
        p \cdot q = q & \implies & \parityOf{p-1} + \parityOf{q} \geq w \\
        \multicolumn{3}{l}{
            \parityOf{p \cdot q} = \min(w, \parityOf{p} + \parityOf{q})
        } \\
    \end{array}%
    \]%
\end{lemma}

\subsection{Incremental Linearization\label{sec:incremental-linearization}}

\polysat{} includes incremental linearization rules
for the cases where variables are $0$, $1$, $-1$, or powers of two.
Note that our vocabulary of incremental linearization lemmas
is considerably smaller than what is used for non-linear integer
arithmetic~\cite{DBLP:conf/sat/CimattiGIRS18},
but it is also materially different as it operates over modular semantics of bit-vector operations.
Notably, we do not include here inferences for deriving ordering constraints,
such as $a > b \land c > 0 \implies ac > bc$, which holds for integers, but not for bit-vectors.
Note that Lemma~\ref{lemma:saturation-modulo-multiplication} includes ordering constraints,
but only for the cases where relevant uses of multiplication do not overflow.

\begin{lemma}[Incremental Linearization]
    \[
        \begin{array}{lcl}
            p = 0 & \implies & p \cdot q = 0 \\
            p = 1 & \implies & p \cdot q = q \\
            p = -1 & \implies & p \cdot q = -q \\
            p = 2^k & \implies & p \cdot q = 2^k q \quad (k = 1, \ldots, w - 1) \\
            p \cdot q = 1  & \implies & p = 1 \lor \overmul{p}{q} \\
            p \cdot q = q  & \implies & p = 1 \lor q = 0 \lor \overmul{p}{q} \\
        \end{array}
    \]
\end{lemma}

\subsection{Bit-blasting Rules\label{sec:bit-blasting}}

As a final resort, \polysat{} admits bit-blasting.
A product $x := p \cdot q$ can be equivalently
represented as $\sum_i 2^i \bit{p}{i} q$.

The other primitive operations (bit-wise \emph{and}, bit-wise \emph{or}, left shift, logical and arithmetic right shift)
are unfolded using blasting as follows.

\begin{lemma}[$x \coloneqq \band{p}{q}$]
    \label{lemma:bb-and}
    Bit-wise \emph{and} ``$\bvand$`` is handled using standard axioms,
    that fall back to bit-blasting at each index~$i$
    if the basic algebraic properties hold,
    but $x$ still does not evaluate to the bit-wise \emph{and} of $p, q$.
    \[
    \begin{array}{llll}
        \top            & \implies & x \bvule p \\
        p = 0           & \implies & x = 0 \\
        p = -1          & \implies & x = q \\
        p = q           & \implies & x = p \\
        p[i] \land q[i] & \implies & x[i] & \text{for each $0 \leq i < w$} \\
        x[i]            & \implies & p[i] & \text{for each $0 \leq i < w$} \\
    \end{array}
    \]
    Note that we do not list symmetric rules, e.g., $x \bvule q$.
\end{lemma}

Bit-wise~\emph{or} is handled analogously.
For shift operations, we split on the value of the second argument.
Details may be found in Appendix~\ref{appendix:additional-lemmas}.

\polysat{} also performs partial bit-blasting for multiplication overflow predicates.
It is based on partitioning the conditions for overflow
by using the sum of most significant bits into three cases.
To describe these, first let us define the shorthand~$\Msb{p}$
for the one-based index of the most significant bit of~$p$.
For example, $\Msb{1} = 1, \Msb{2} = 2$.
It can be defined indirectly using the equivalence
$\Msb{p} \geq i \Longleftrightarrow  p \bvuge 2^ {i-1}$ for $1 \leq i \leq w$.
The cases are
\[
    \begin{array}{rcll}
    \Msb{p} + \Msb{q} \geq w + 2 & \implies & \overmul{p}{q} \\
    \Msb{p} + \Msb{q} \leq w & \implies & \neg\overmul{p}{q} \\
    \Msb{p} + \Msb{q} = w + 1 & \implies \\
    \multicolumn{3}{r}{
        (\overmul{p}{q} \Longleftrightarrow (0p) \cdot (0q) \bvuge 2^w),
    }
    \end{array}
\]
where $0p$ and $0q$ stand for the zero-extension by a single bit of $p$ and $q$, respectively.
In other words, when the most significant bits add up to $w$,
multiplication overflow affects exactly one additional bit,
so it suffices to extend~$p$ and~$q$ by a single bit to determine overflow.

\section{Experiments\label{sec:experiments}}

We  evaluated our \polysat{} prototype\footnote{%
    Available at \url{https://github.com/Z3Prover/z3/tree/poly}.
    This paper refers to commit \texttt{16fb86b636047fd79ad5827f768b6f26d8812948}.
    To select \polysat{} for bit-vector solving,
    add the following options: \texttt{sat.smt=true tactic.default\_tactic=smt smt.bv.solver=1}.
}
against recent versions of several state-of-the-art
SMT solvers
on the following four benchmark sets:
the category \texttt{QF\_BV} from SMT-LIB~\cite{SMTLIB} (release 2023, non-incremental);
the BV2SMV benchmarks featuring large bit-widths~\cite{BV2SMV};
14 benchmarks from smart contract verification related to the Certora prover~\cite{DBLP:journals/pacmpl/AlbertGRRRS20};
and a set of benchmarks from the Alive2 compiler verification project~\cite{Alive2}.
Note that the STP solver~\cite{STP} does not support the logic \texttt{QF\_UFBV}
used by some of the Certora benchmarks.

Our experiments were performed on 
a TU Wien cluster, 
where each compute node contains two AMD Epyc 7502 processors,
each of which has 32~CPU cores running at 2.5\,GHz.
Each compute node is equipped with 1008\,GiB of physical memory
that is split into eight memory nodes of 126\,GiB each,
with eight logical CPUs assigned to each node.
We used \texttt{runexec} from the
benchmarking framework \textsc{BenchExec}~\cite{BeyerLoweWendler:2017:benchexec}
to assign each benchmark process to a different CPU core and its corresponding memory node,
aiming to balance the load across memory nodes.
Further, we used \textsc{GNU Parallel}~\cite{tange_2024_10558745}
to schedule benchmark processes in parallel.

Our results are summarized in \cref{tab:experiments}
and indicate that there is no clear winner among existing solvers,
especially when focusing on symbolic reasoning over bit-vectors (without bit-blasting).
Overall, the performance of \polysat{} is comparable to the other word-level approaches
on the BV2SMV benchmark set, however in general, more work is needed.
Importantly, \polysat{} complements \zthree{} with word-level bit-vector reasoning.
In particular, our experimental analysis found that~\polysat{}
solved 135 problems that~\zthree{} did not solve
and 404 problems that \zthree{}-IntBlast did not solve
(40 of which neither \zthree{} nor \zthree{}-IntBlast solved).
Further combinations of complementary approaches of word-level reasoning with bit-blasting
is a promising directions to explore.

\tabcolsep1mm
\begin{table}[t]
    \centering
    \newcommand{\red}[1]{{\textcolor{red}{#1}}}
\begin{tabular}{ll|rr|rr|rr|rr}
& & \multicolumn{2}{c|}{SMT-LIB} & \multicolumn{2}{c|}{BV2SMV} & \multicolumn{2}{c|}{\makecell{Smart\\Contracts}} & \multicolumn{2}{c}{Alive2}  \\
& & sat & unsat & sat & unsat & sat & unsat & sat & unsat  \\
\hline\Tstrut
\multirow{5}{*}{\begin{turn}{90}Bit-blasting\end{turn}}
& \bitwuzla{} \cite{Bitwuzla}                   & 17\,745 & 27\,203         & 32 & 115      & 1 & 3         & 39 & 3\,954 \\
& \cvcfive{} \cite{cvc5}                        & 16\,417 & 25\,922         & 31 & 114      & 0 & 4         & 39 & 2\,722 \\
& \stp{} \cite{STP}                             & 17\,462 & 27\,011         & 24 & 115      & - & -         & 39 & 2\,893 \\
& \yicestwo{} \cite{Yices2}                     & 17\,589 & 26\,600         & 24 & 107      & 0 & 3         & 39 & 1\,519 \\
& \zthree{} \cite{Z3}                           & 16\,112 & 25\,597         & 29 &  94      & 0 & 3         & 39 & 1\,514 \\
\hline\Tstrut
\multirow{3}{*}{\begin{turn}{90}Word-lvl\,\,\end{turn}}
& \cvcfive{}-IntBlast \cite{IntBlasting}        & 11\,251 & 24\,376         & 32 &  64      & 1 & 9         &  5 & 1\,047 \\
& \yicestwo{}-mcsat  \cite{BitvectorsMCSAT}     & 14\,155 & 22\,396         & 24 & 101      & 1 & 4         & 23 & 2\,562 \\
& \zthree{}-IntBlast                            & 10\,912 & 24\,371         & 28 &  56      & 1 & 5         & 30 &    921 \\
& \zthree{}-\polysat{}                          &  7\,297 & 20\,080         & 28 &  63      & 0 & 3         &  0 &     21 \\
\hline\Tstrut
& Total & \multicolumn{2}{c|}{46\,191} & \multicolumn{2}{c|}{192} & \multicolumn{2}{c|}{14} & \multicolumn{2}{c}{12\,951}  \\
\end{tabular}

    \caption{%
        Number of problems solved within~60\,s
        for several benchmark sets.
        The upper five solvers are based on bit-blasting,
        while the lower four solvers use word-level techniques.
    }
    \label{tab:experiments}
\end{table}

\nocite{Bitwuzla}
\nocite{cvc5}
\nocite{STP}
\nocite{Yices2}
\nocite{Z3}
\nocite{IntBlasting}
\nocite{BitvectorsMCSAT}
\nocite{SMTLIB}
\nocite{BV2SMV}
\nocite{Alive2}
\nocite{tange_2024_10558745}

\section{Conclusion\label{sec:conclusion}}

We introduced \polysat{}, a general purpose word-level bit-vector solver,
to overcome the scalability issue of bit-blasting over large bit-vectors.
\polysat{} integrates into CDCL(T)-based SMT solving,
generalizes interval-based reasoning,
and performs incremental linearization of constraints.
\polysat{} is implemented in
the SMT solver~\zthree{} and complements bit-vector reasoning in~\zthree{}.

\noindent\textbf{Acknowledgements.}
We thank Mooly Sagiv and Alexander Nutz for thorough discussions on \polysat{} applications.
This work was partially supported by
the ERC Consolidator Grant ARTIST 101002685,
the TU Wien Doctoral College SecInt,
the FWF SFB project SpyCoDe F8504, the FWF ESPRIT grant 10.55776/ESP666,
and the Amazon Research Award 2023 QuAT.

\bibliographystyle{abbrv} 
\IEEEtriggeratref{16}
\bibliography{refs}

\newpage
\appendix
\subsection{Additional Lemmas\label{appendix:additional-lemmas}}

\begin{lemma}
\label{lemma:saturation-modulo-multiplication-full}
Extended version of Lemma~\ref{lemma:saturation-modulo-multiplication}.
\[
\begin{array}{llllllllll}
p x \bvult q x                  &\implies&& p \neq q                                \\
p x \bvult q x                  &\implies&& \overmul{p}{x}      &\lor& p \bvult q   \\
p x \bvult q x                  &\implies&& \overmul{-q}{x}     &\lor& p \bvult q   \\
p x \bvult q x                  &\implies&& \overmul{q}{-x}     &\lor& p \bvugt q   \\
~                               &&\lor&     p = 0                                   \\
p x \bvult q x                  &\implies&& \overmul{-p}{-x}    &\lor& p \bvugt q   \\
~                               &&\lor&     p = 0                                   \\
p x \bvule q x                  &\implies&& \overmul{p}{x}      &\lor& p \bvule q   \\
~                               &&\lor&     x = 0                                   \\
p x \bvule q x                  &\implies&& \overmul{-q}{x}     &\lor& p \bvule q   \\
~                               &&\lor&     x = 0               &\lor& q = 0        \\
p x \bvule q x                  &\implies&& \overmul{q}{-x}     &\lor& p \bvuge q   \\
~                               &&\lor&     x = 0               &\lor& p = 0        \\
p x \bvule q x                  &\implies&& \overmul{-p}{-x}    &\lor& p \bvuge q   \\
~                               &&\lor&     x = 0               &\lor& p = 0        \\
px + s \bvule q &\implies& & \overmul{p}{x}   &\lor & \overadd{px}{s}  \\
&                       & \lor & p r \bvule q &\lor & x \bvult r       \\
p \bvule x \land qx \bvule r    &\implies&& \overmul{q}{x} & \lor& pq \bvule r  \\
p \bvule x \land qx \bvult r    &\implies&& \overmul{q}{x} & \lor& pq \bvult r  \\
p \bvult x \land qx \bvule r    &\implies&& \overmul{q}{x} & \lor& pq \bvult r  \\
~                               &&\lor&     q = 0                               \\
p \bvult x \land qx \bvule r    &\implies&& \overmul{q}{x} & \lor& pq \bvult r  \\
~                               &&\lor&     r = 0                               \\
p \bvule qx \land x \bvule r    &\implies&& \overmul{q}{r} & \lor& p \bvule qr  \\
p \bvult qx \land x \bvule r    &\implies&& \overmul{q}{r} & \lor& p \bvult qr  \\
p \bvule qx \land x \bvult r    &\implies&& \overmul{q}{r} & \lor& p \bvult qr  \\
~                               &&\lor&     p = 0                               \\
p \bvule qx \land x \bvult r    &\implies&& \overmul{q}{r} & \lor& p \bvult qr  \\
~                               &&\lor&     q = 0                               \\
\end{array}
\]
Note that these rules do not require $x \not\in p, q, r, s$,
so they can be applied even when the degree of $x$ is larger than $1$.
\end{lemma}

\begin{lemma}[$x \coloneqq \bor{p}{q}$]
    \label{lemma:bb-or}
    Bit-wise \emph{or} is handled similarly as bit-wise~\emph{and}.
    \[
    \begin{array}{llll}
        \top            & \implies & x \bvuge p \\
        p = 0           & \implies & x = q  \\
        p = -1          & \implies & x = -1 \\
        p = q           & \implies & x = p  \\
        p[i]            & \implies & x[i]           & \text{for each $0 \leq i < w$} \\
        x[i]            & \implies & p[i] \lor q[i] & \text{for each $0 \leq i < w$} \\
    \end{array}
    \]
\end{lemma}

\begin{lemma}[$x \coloneqq \bshl{p}{q}$]
    \label{lemma:bb-shl}
    For shift operations,
    we split on the second argument.
    \[
    \begin{array}{lll}
        q \bvuge w  & \implies & x = 0 \\
        q = 0       & \implies & x = p \\
        q = i       & \implies & x = 2^{i}p \\
    \end{array}
    \]
    for all constants $i$ such that $0 < i < w$.
\end{lemma}

\begin{lemma}[$x \coloneqq \blshr{p}{q}$]
    \label{lemma:bb-lshr}
    Logical right-shift is analogous.
    \[
    \begin{array}{lll}
        q \bvuge w  & \implies & x = 0 \\
        q = 0       & \implies & x = p \\
        q = i       & \implies & 2^ix \bvule p \bvule 2^ix + 2^i - 1 \land x \bvult 2^{w-i} \\
    \end{array}
    \]
    for all constants $i$ such that $0 < i < w$.
\end{lemma}

\begin{lemma}[$x \coloneqq \bashr{p}{q}$]
    \label{lemma:bb-ashr}
    The arithmetic right-shift must take into account the sign bit $\bit{p}{w-1}$.
    \[
    \begin{array}{lll}
        \bit{p}{w-1}        \land q \bvuge w    & \implies & x = -1 \\
        \lnot \bit{p}{w-1}  \land q \bvuge w    & \implies & x = 0 \\
                                  q \bvuge w    & \implies & x + 1 \bvule 1 \\
        q = 0                                   & \implies & x = p \\
        q = i                                   & \implies & 2^ix \bvule p \bvule 2^ix + 2^i - 1 \\
        \bit{p}{w-1}        \land q = i         & \implies & x \bvuge 2^w - 2^{w-i-1} \\
        \lnot \bit{p}{w-1}  \land q = i         & \implies & x \bvult 2^{w-i-1} \\
    \end{array}
    \]
    for all constants $i$ such that $0 < i < w$.
\end{lemma}

\end{document}